# Attention Patterns Detection using Brain Computer Interfaces

Felix G. Hamza-Lup
Georgia Southern University
Savannah, GA, USA
fhamzalup@GeorgiaSouthern.edu

Aditya Suri
Georgia Southern University
Statesboro, GA, USA
as14124@GeorgiaSouthern.edu

Ionut E. Iacob
Georgia Southern University
Statesboro, GA, USA
ieiacob@GeorgiaSouthern.edu

Ioana R. Goldbach
Valahia University of Targoviste
Targoviste, DB, Romania
ioana.goldbach@icstm.ro

Lateef Rasheed
Georgia Southern University
Statesboro, GA, USA
lr10486@GeorgiaSouthern.edu

Paul N. Borza
Transilvania University of Brasov
Brasov BV, Romania
borzapn@UnitBV.ro

## ABSTRACT

The human brain provides a range of functions such as expressing emotions, controlling the rate of breathing, etc., and its study has attracted the interest of scientists for many years. As machine learning models become more sophisticated, and biometric data becomes more readily available through new non-invasive technologies, it becomes increasingly possible to gain access to interesting biometric data that could revolutionize Human-Computer Interaction. In this research, we propose a method to assess and quantify human attention levels and their effects on learning. In our study, we employ a brain computer interface (BCI) capable of detecting brain wave activity and displaying the corresponding electroencephalograms (EEG). We train recurrent neural networks (RNNS) to identify the type of activity an individual is performing.

## 1 INTRODUCTION

Brain activity can be tested through applicable devices that measure wavelengths such as an electroencephalograph (EEG). An EEG is important as it enables the recording of signals that are sent out by the brain on several frequencies. Different emotions show different characteristics in a brainwave. For instance, when someone is tired or slow, the brainwave amplitude may change compared to someone who is more alert, who is showing hyperactive brainwaves. There are different classifications for identifying the type of brainwaves such as gamma, beta, alpha, theta, and delta. Alpha waves are generated when someone is in meditation or is learning something. This type of wave is the most prominent in an EEG with a frequency that ranges from 8 to 13 Hz. On the other hand, beta waves have a lower amplitude with a frequency range of 13 to 22 Hz. They change when an individual is excited or focusing on something. Theta waves have frequencies that range from 4 to 8 Hz, which are considered low frequencies. This type of wave is shown more when an individual is experiencing deep meditation. On the contrary, delta waves have the lowest frequencies that range from 1 to 4 Hz. Lastly, gamma waves have a very high-frequency range, which is above 32 Hz. This wave mainly helps in identifying the effects of the external world on the neural network [3].

## 2 SYSTEM DESIGN AND EXPERIMENTAL RESULTS

For our analysis, we have used the Neuro-Marketing data [3], which disseminates between *Like/Dislike* mind state of 25 subjects who were observing several images. Multiple time series (signals) were collected from each subject, using a more advanced device (Emotiv EPOC+) capable of collecting 14 brain signals simultaneously. For this study, we have selected all signals produced by one sensor placed on the subjects' forehead. Our choice was consistent with designing our data collection experiments using a Neurosky™ Mindwave Mobile 2 device capable of collecting a single signal, typically from the forehead area. Overall, in this analysis, we have used 1045 signals collected using the forehead sensor. Figure 1 (left) shows two such signals, corresponding to *Like/Dislike* signals, respectively. A quick observation that the *Dislike* signals have more variance (some related work relies on such observation) proves insufficient for performing an accurate analysis.

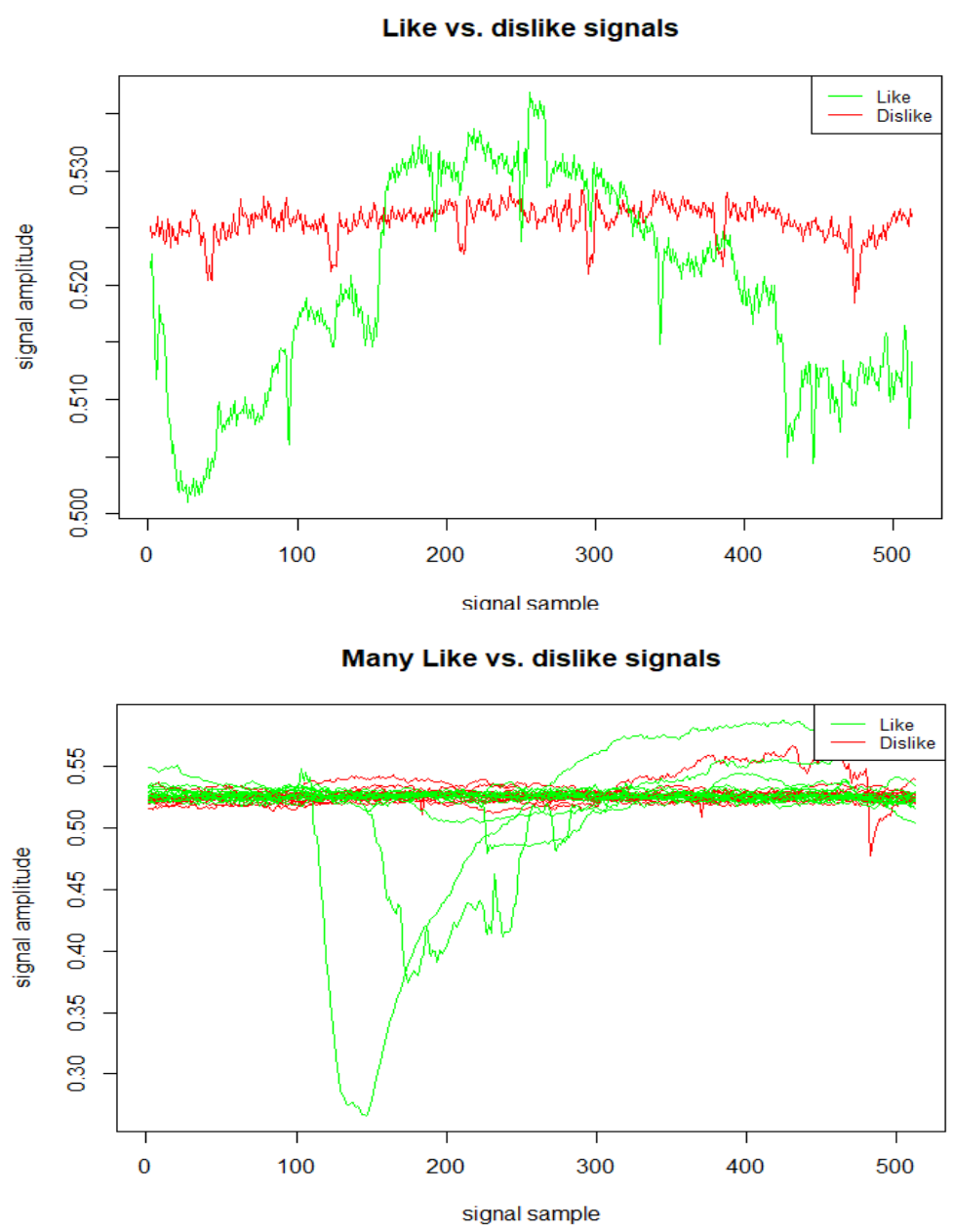

**Figure 1: Like/Dislike Sample Signals (Top); Multiple Like/Dislike Signals (Bottom)**

Figure 1 (right) shows multiple signals *Like/Dislike* plotted together and once can see that only some *Dislike* signals have more variance than the *Like* signals. Instead, we have normalized the signals and extracted the signal's lower frequency components using spline filters, then used an Artificial Neural Network (ANN) model to perform classification on these components.

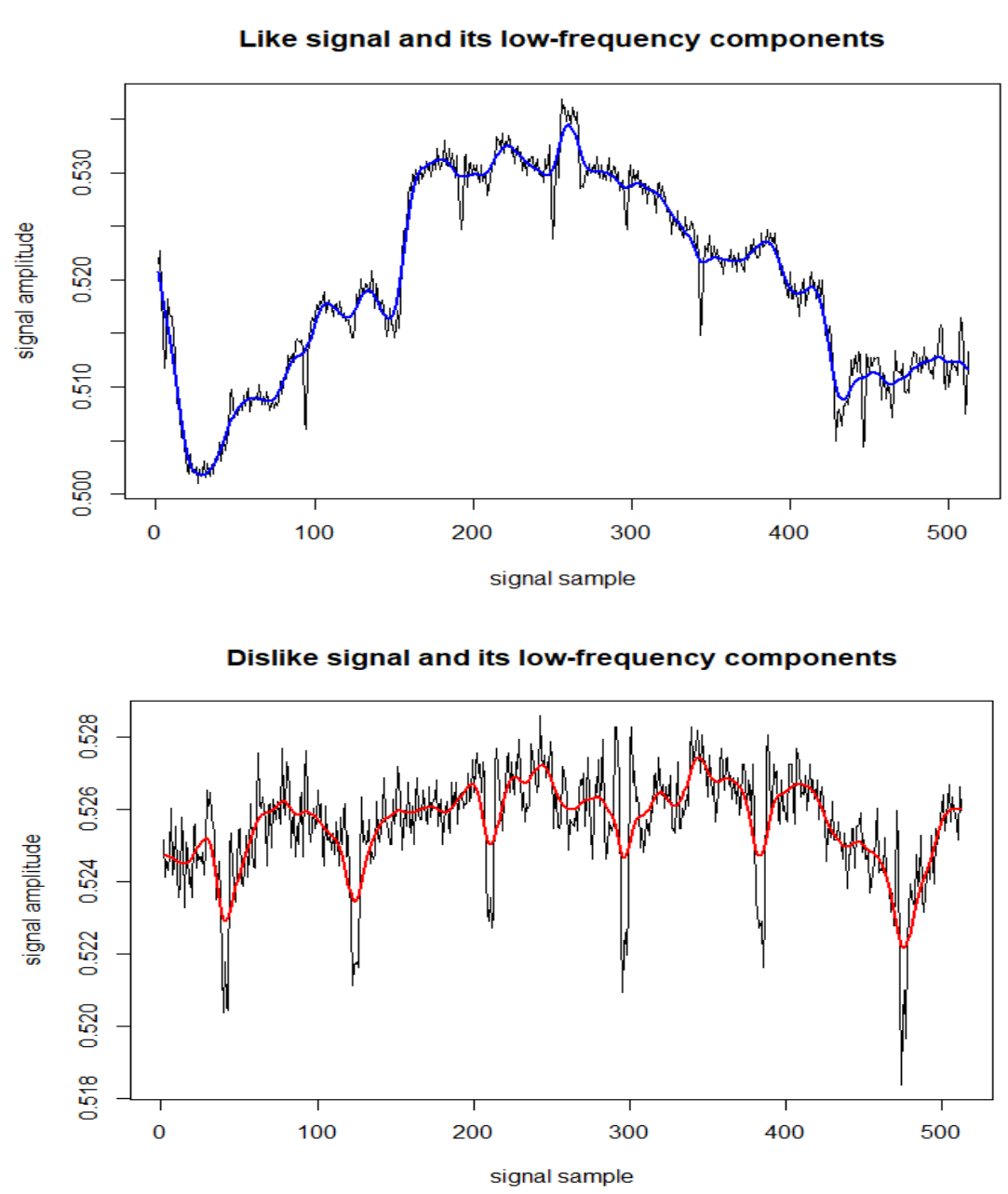

**Figure 2: Sample Like and Dislike Signals and Their Low-Frequency Components**

Figure 2 shows sample *Like/Dislike* signals with their low-frequency components, respectively. For these experiments, we used an ANN model created with Keras and R to perform the binary classification of signals' low-frequency components. Unsurprisingly, as also reported in [3], the results were encouraging but rather limited in terms of accuracy (about 60%). The breakthrough happened when we used a non-linear transformation on the un-normalized low-frequency data coupled with bootstrapping. The ANN model was created of two hidden layers with 128 and 32 hidden neurons, respectively, and with "relu" activation functions for each layer. As reported in Figure 3 the accuracy of the model for processing both validation and training data jumped to the upper 90%.

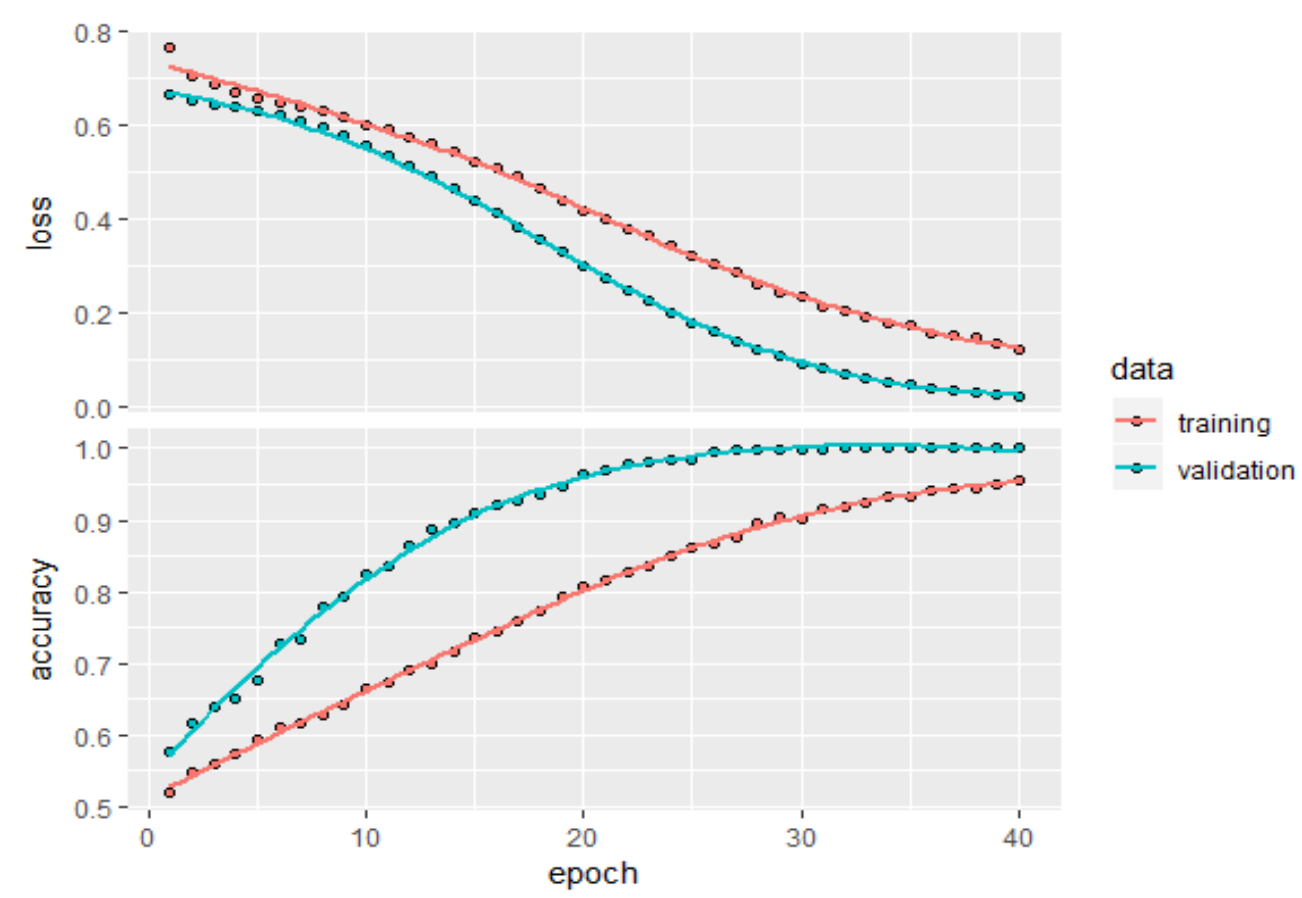

**Figure 3: ANN Model Classification Results**

## 3 CONCLUSION

The brain is a vital organ that transmits electrical impulses (brainwaves) which, can be measured via tools such as Emotiv EPOC+ or Neurosky™ Mindwave devices. The types of activity an individual is performing or the emotion they are feeling change the frequency of the brainwave [1, 2]. The goal of this study was to test the potential of ANN models to measure brain activity, which will potentially help in building tools that will aid in detecting/controlling human emotions. Experimental results were extremely encouraging. We performed filtering on brainwave collected through a single sensor, non-linear transformation of the data, and bootstrapping to produce high-accuracy classification using an ANN model. Our results outperform previous classification attempts [3] on the same dataset.